\documentstyle[12pt]{article}
\newcommand{\eqr}[1]{(\ref{#1})}
\newfont{\feff}{cmti10}
\def\undertext#1{\vtop{\hbox{#1}\kern 1pt \hrule}}
\def\ra{\rightarrow}
\def\s{\sigma}
\def\a{\alpha}
\def\m{m_0^2}
\def\Px{\Phi_x}
\def\Py{\Phi_y}
\def\la{\lambda_A}

\def\d{\hbox{dim}\,}
\def\mod{\hbox{mod}\,}
\def\mN{~~(\mod N)}
\newcommand{\tr}[1]{\,{\rm tr}\,#1\,}
\def\e{{\,\rm e}\,}
\def\be{\begin{equation}}
\def\ee{\end{equation}}
\def\bea{\begin{eqnarray}}
\def\eea{\end{eqnarray}}
\def\eqref#1{\ref{(#1)}}
\begin{document}
\begin{titlepage}
\begin{flushright} TAUP 1996-92\\ September 1992 \end{flushright}
\vskip2cm
\centerline{\LARGE \bf From hermitian matrix model}
\vskip.3cm
\centerline{\LARGE \bf to lattice gauge theory}
\vskip1.5cm
\centerline{\large B.Rusakov}
\vskip.5cm
\begin{center}
\it School of Physics and Astronomy \\
\it Raymond and Beverly Sacler Faculty of Exact Sciences \\
\it Tel-Aviv University, Israel \\
\end{center}
\vskip2cm
\begin{abstract}
I consider a lattice model of a gauge field interacting with
matrix-valued scalars in $D$ dimensions. The model includes an
adjustable parameter $\s$, which plays role of the string tension.
In the limit $\s=\infty$ the model coincides with Kazakov-Migdal's
``induced QCD", where Wilson loops obey a zero area law. The limit
$\s=0$, where Wilson loops $W(C)=1$ independently of the size of
the loop, corresponds to the Hermitian matrix model. For $D=2$ and
$D=3$ I show that the model obeys the same combinatorics as the standard
LGT and therefore one may expect the area law behavior.
In the strong coupling expansion such a behavior is demonstrated.

\noindent
\end{abstract}

\end{titlepage}

\section{The model.}

In what follows I propose a lattice gauge model interpolating
between the Hermitian matrix model \cite{MatrMod}
and Kazakov-Migdal's ``induced QCD" \cite{KM}. The model is very
close to standard Wilson LGT.
For $D=2$ and $D=3$ I show that the model obeys the same
combinatorics as the standard LGT and, therefore, if we generally expect
area law in the latter then one may expect it in the former.
Within the strong coupling expansion the area law in
the model is demonstrated below.
The reason for the combinatorial similarity of our model to the
standard LGT is the $Z_N$ symmetry.

Let us start from the Hermitian matrix model
\cite{MatrMod} in $D$ dimensions
\be
Z= \int \prod_{x}D\Px \e^{\tr V(\Px)}
\prod_{<xy>} \e^{N\tr[\Px\Py]} .
\label{Z}
\ee
After diagonalizing the hermitian $N\times N$ matrices $\Px$
by $SU(N)$ matrices $U_x: ~ \Px\ra U_x\Px U_x^\dagger$ and defining
for each link $<xy>$ of the $D$-dimensional lattice the
$SU(N)$ variables $U_{xy}=U_x^\dagger U_y$, eq.\eqr{Z} reads
\be
Z= \int \prod_{x} d\Px \Delta^2 (\Px)
\e^{ \tr V(\Px)}
\int\prod_{<xy>}DU_{xy}
\e^{N\tr[\Px U_{xy}\Py U_{xy}^\dagger]}
\prod_{f}\delta (I,U_f).
\label{Zre}
\ee
where $\Px$'s are now {\it diagonal} matrices, $U_f$ is the ordered
product of $U_{xy}$'s along the face (fundamental polygon) $f$
\footnote{In the case of the square lattice,
$f$ is the fundamental plaquette.}
and $\Delta (\Px)$ is the Vandermonde
determinant. The gauge invariant $\delta$-function is equivalent to
a non-po\-ly\-no\-mial gauge self-interaction: by definition,
the $\delta$-function is a sum over irreducible representations $r$,
\be
\delta(I,U_f)=\sum_{r}\d_r\chi_r(U_f),
\label{delta}
\ee
where $\chi_r(U)$ and $\d_r=\chi_r(I)$ are characters and dimensions of
the $r$'s. Using the Cauchy formula, eq.\eqr{delta} can be expressed as
\be
\sum_{r}\d_r\chi_r(U_f)=\prod_{k=1}^{N}[1-(U_f)_{kk}]^{-N}=
\exp \left(-N\tr\log(I-U_f)\right)   .
\label{deltare}
\ee
Hence, the $U$-dependent part of action takes the form
\be
S=N\sum_{<xy>}\tr[\Px U_{xy}\Py U_{xy}^\dagger]
+N\sum_{f}\sum_{q=1}^{\infty}~{1\over q} \tr (U_f)^q .
\label{Act}
\ee

The first term in \eqr{Act} coincides with the action of the
``induced QCD"
\cite{KM}. The properties of the model \cite{KM} have been investigated
intensively \cite{KM}-\cite{KKMM} and it has been observed that
the local $Z_N$
symmetry of the Kazakov-Migdal action results in the so called ``strong
confinement", or ``zero area law" (see Ref.\cite{Ko} for details): the
Wilson loop $W(C)$ vanishes unless the spanned
area $A(C)$ equals zero:
\be
W(C)=0   \,,\qquad A(C)\neq 0 .
\label{area0}
\ee
If the (minimal) area law $W(C)\sim\exp(-\s A(C))$ is presumed, then the
behavior \eqr{area0} corresponds to an infinite string tension,
$\s=\infty$.

We now make the simple observation that the second term in \eqr{Act},
arising from $\delta$-function \eqr{delta}, {\it breaks} local $Z_N$
and leads to ``deconfinement"
\be
W(C)=1  \,,\qquad A(C)\neq 0  ,
\label{de}
\ee
independently of the size of the loop $C$. Indeed, any product of
$U_{xy}$'s along a contour $C$ is equal to product of $U_f$'s, where
faces $f$'s fill $C$. Each such $U_f=I$, due to the $\delta$-function
and, hence, $\prod_{<xy>\in C} U_{xy}=I$. It is also clear if one
recalls that the matrix model \eqr{Z} is actually independent of
the $U$'s.

Formally, this corresponds to zero string tension, $\s=0$.

It is natural to try to generalize this model to one
interpolating between the two limiting cases \eqr{area0} and \eqr{de}.

Consider the following simple extension:
\be
\tr\log(I-U_f) ~ \ra ~ \tr\log(I-\a U_f)
\label{ext}
\ee
where $0\leq\a\leq 1$, with $\a=\e^{-\s}$. Thus $0\leq\s\leq\infty$, and
$\s=\infty$ corresponds to the Kazakov-Migdal ``strong confinement"
\eqr{area0}, while $\s=0$ corresponds to deconfinement \eqr{de}.

The action now reads
\be
S=N\sum_{<xy>}\tr[\Px U_{xy}\Py U_{xy}^\dagger]
+N\sum_{f}\sum_{q=1}^{\infty}~{{\a^q}\over q} \tr (U_f)^q    .
\label{model}
\ee

This implies that instead of \eqr{delta} for each face $f$,
we have a factor
\be
\sum_{r}\a^{\nu_r}\d_r\chi_r(U_f)
\label{deltaa}
\ee
where $\nu_r$ is the sum of all components of highest weight of an
irreducible representation $r\equiv\left\{n_1,\dots,n_N\right\}$:
\be
\nu_r=\sum_{k=1}^{N} n_k .
\label{nu}\ee

\section{Loop averages.}

If we expect an area law in the standard LGT, then
it is natural to expect this property in the model \eqr{model}.
Naively, we have checked this already in the two limits,
\eqr{area0} and \eqr{de} corresponding to $\s=\infty$ and $\s=0$
respectively. Now I demonstrate for $D=2$ and $D=3$ that the model
with arbitrary $0<\s<\infty$ obeys the same
combinatorics as the standard LGT and that for large $\s$ the
model obeys an area law.

Remarkably, a crucial role is again played by the invariance of the term
$\tr[\Px U_{xy}\Py U_{xy}^\dagger]$ with respect to $Z_N$
transformations
\footnote{In fact, there is also invariance with respect
to arbitrary diagonal $SU(N)$-left and -right rotations.}
\be
U ~ \ra ~ Z_N U \, \qquad \left(U ~ \ra ~ U Z_N\right) .
\label{rot}
\ee

\subsection{$D=2.$}

To calculate loop averages and partition functions
we need to average the product
(over all faces) of the quantities \eqr{deltaa} with respect to the
Kazakov-Migdal action. The invariance of this action and of the Haar
measure under transformations \eqr{rot}
immediately gives selection rules for one-link integrals:

\noindent
-- for free links
(only such links contribute to the partition function):
\be
\int DU \e^{N\tr[\Px U\Py U^\dagger]}
\chi_{r_1}(UV_1)\chi_{r_2}(U^{\dagger}V_2)
 ~ \sim ~ \delta_{\nu_1,\nu_2} ~ \mN
\label{sel0}
\ee
-- while for links belonging to a Wilson loop:
\be
\int DU \e^{N\tr[\Px U\Py U^\dagger]}
\chi_{r_1}(UV_1)\chi_{r_2}(U^{\dagger}V_2) \tr(UV)
 ~ \sim ~ \delta_{\nu_1+1,\nu_2} ~ \mN
\label{sel1}
\ee

These selection rules provide all the combinatorics and, in particular,
allow us to calculate Wilson loops at strong coupling
(large $\s$). The calculation in this case reduces to counting
powers of $\a$ and to selecting the lowest order terms in the same way
as it can be done in the standard Wilson lattice $SU(N)$ gauge theory.
For example, for a simple loop with the topology of a circle the
lowest order non-zero term corresponds to the trivial
representation for each face outside the contour and to the
fundamental representation inside. Then, the result
is that the total lowest power is equal to
the number of faces $A(C)$ enclosed by loop $C$.
Thus, we obtain the area law,
\be
W(C) ~ \sim ~ \e^{-\s A(C)}
\label{Loop}
\ee
up to a factor independent of $\s$.

It is worthwhile to remark here that for
$D=2$ the model becomes exactly solvable in the limit of infinite
mass scalars. Indeed, following \cite{KM}, consider a quadratic potential
of the scalar field:
\be \tr V(\Px)=N\m\tr (\Px^2) \ee
After gaussian integration over $\Phi$'s we have the Kazakov-Migdal
action of ``induced QCD", which is proportional to negative powers of
$\m$. Hence, in the limit $\m=\infty$, the gauge field is completely
decoupled from the scalars and instead of \eqr{sel0} we have an
orthogonality condition for characters
\be
\int DU\chi_{r_1}(UV_1)\chi_{r_2}(U^{\dagger}V_2)
= \delta_{r_1,r_2}{{\chi_{r_1}(V_1V_2)}\over{\d_{r_1}}}
\label{ort0}
\ee
and instead of \eqr{sel1} we have the Wigner coefficient
\be
\int DU \chi_{r_1}(U)\chi_{r_2}(U^{\dagger}) \tr(U)
= D_{r_1,r_2}^f ~ ~ ,
\label{Wig}
\ee
which makes the model to be exactly solvable.
The subsequent calculations follow the approach suggested in \cite{Mig}.
(Full combinatorial details, as well as expressions for Wilson loops and
partition function on the most general Riemann surface are explicitly
given in \cite{Ru}.) In
particular, the result for a simple Wilson loop \eqr{Loop} becomes exact
for an arbitrary $\s$: $W(C)=\exp(-\s A(C))$.

\subsection{$ D=3.$}

In this case, the symmetry \eqr{rot} leads to the
following selection rules:

\noindent
-- for free links:
\bea
\int DU \e^{N\tr[\Px U\Py U^\dagger]}
\chi_{r_1}(UV_1)\chi_{r_2}(U^{\dagger}V_2)
\chi_{r_3}(UV_3)\chi_{r_4}(U^{\dagger}V_4) ~ \sim
\nonumber \eea
\be \sim ~ \delta_{\nu_1+\nu_3,\nu_2+\nu_4} ~ \mN
\label{sel2}
\ee
-- and for links belonging to a Wilson loop:
\bea
\int DU \e^{N\tr[\Px U\Py U^\dagger]}
\chi_{r_1}(UV_1)\chi_{r_2}(U^{\dagger}V_2)
\chi_{r_3}(UV_3)\chi_{r_4}(U^{\dagger}V_4) \tr(UV) ~ \sim
\nonumber \eea
\be \sim ~ \delta_{\nu_1+\nu_3+1,\nu_2+\nu_4} ~ \mN
\label{sel3}
\ee

The strategy of further calculation of the Wilson loop in the strong
coupling expansion is straightforward and closely follows
the $D=2$ case, even though the
combinatorial details are somewhat more complicated.
As a result, we again obtain the area-like behavior,
\be W(C) ~ \sim ~ \sum_{A(C):\partial A=C}\la \e^{-\s A(C)} .
\label{loop}\ee
where $\la$'s are coefficients growing slower than $\exp(\s A(C))$
\footnote{I do not discuss here the problems
related to the competition of coefficients $\la$'s with
$\e^{-\s A(C)}$ in the vicinity of RG-stable point, etc.
My only purpose here is to demonstrate the usual
LGT behavior for the model \eqr{model}.} .
The remark similar to $D=2$ case can be done: in the $\m=\infty$
limit the expressions \eqr{sel2} and \eqr{sel3} become exactly the same
as in the pure Wilson LGT.

\section{Conclusions.}

We have a lattice model of a gauge field interacting with
matrix-valued scalars in $D$ dimensions. The model includes an
adjustable parameter $\s$, which plays role of the string tension.
In the limit $\s=\infty$ the model coincides with Kazakov-Migdal's
``induced QCD", where Wilson loops obey a zero area law. The limit
$\s=0$, where Wilson loops $W(C)=1$ independently of the size of the
loop, corresponds to the Hermitian matrix model. For $D=2$ and $D=3$
we have shown that the model obeys the same combinatorics as the standard
LGT and, therefore, one may expect the area law behavior.
We have demonstrated this property at the
strong coupling. In the massive limit, $\m=\infty$, where the gauge
field is completely decoupled from the scalars the model coincides
(up to inessential details) with the pure Wilson LGT.

Thus, the Hermitian matrix model and the ``induced QCD" can be
considered as the limiting cases of one lattice gauge model.

In order to better understand how the incorporation of the $\a$
term changes the matrix model, it is
fruitful to look at the $D=1$ case (the matrix chain).
In this case the evolution from $\a=1$ to $\a=0$
is very similar to evolution from the closed to the open matrix chain,
which in turn can be described by the action
\be
S=N\sum_{k=1}^{M-1}\tr[\Phi_k\Phi_{k+1}]
+ \a N\tr[\Phi_M\Phi_1]
\label{chain}\ee
where
$M$ is the number of sites of the chain, $\Phi$'s are non-diagonalized.
However, while in \eqr{chain} $\a=0$ corresponds to the chain
being split, it is not the case in our model.
The similarity between our model and the model \eqr{chain}
is the absence of the non-singlet $SU(N)$ states in the $\a=0$ limit
(the ``splitting" point).
Obviously, the ``splitting" point,
$\a=0$, is a singular one. In the language of
LGT, this point corresponds to restoration of local $Z_N$,
resulting in the ``strong confinement".

The unbroken local $Z_N$ symmetry makes
it possible to apply the It\-zyk\-son-Zuber formula
in the Kazakov-Migdal model \cite{KM}
to integrate out the $U$'s reducing the problem
to a master field, etc. As a price for solvability,
the model \cite{KM} obeys an undesirable ``strong confinement".
In our model the gauge self-interaction, $\tr\log(I-U_f)$,
breaks the local $Z_N$, but now
we cannot integrate out the $U$'s exactly.
\pagebreak

\noindent
{\bf Acknowledgements.}
\bigskip

\noindent
I am grateful to
Marek Karliner for discussion and for careful reading of the manuscript.
I thank also Dmitri Boulatov for conversations and for critical remarks.
This research has been supported in part by the Basic Research Foundation
administered by the Israel Academy of Sciences and Humanities, by a
grant from the United States-Israel Binational Science Foundation (BSF),
Jerusalem, Israel, and also by the Israel Ministry of Absorption.

\end{document}